\documentclass[a4paper,11pt]{article}
\pdfoutput=1 % if your are submitting a pdflatex (i.e. if you have
\usepackage{jheppub, placeins, mathtools,notes2bib } % for details on the use of the package, please
\usepackage[T1]{fontenc} % if needed

\title{\boldmath Canonical analysis of E$_{6(6)}(\mathbb{R})$ invariant\\ five dimensional (super-)gravity}

\author[]{Lars T. Kreutzer}%,\note{Corresponding author.}}

\affiliation[]{Max Planck Institute for Gravitational Physics (Albert Einstein Institute), \\Am M\"uhlenberg 1, 14476 Potsdam, Germany}
\affiliation[]{Berlin Mathematical School (BMS), Berlin, Germany}

\emailAdd{lars.kreutzer@aei.mpg.de}

\abstract{We investigate the canonical structure of the bosonic sector of the unique maximal supergravity theory in five dimensions that is manifestly invariant under the global action of E$_{6(6)}(\mathbb{R})$. Starting from the Lagrangian formulation of the theory we construct the Hamiltonian formulation and the full set of canonical constraints. We determine all gauge transformations and compute the algebra formed by the canonical constraints under the Poisson bracket. We re-derive the number of physical degrees of freedom and construct the extended Hamiltonian, describing the most general time evolution of the theory, where the full gauge freedom is manifest. \\
\vspace{8cm}\\
{\footnotesize This article may be downloaded for personal use only. Any other use requires prior permission of the author and AIP Publishing. This article appeared in \textbf{Journal of Mathematical Physics 62, 032302 (2021)} and may be found at \url{https://doi.org/10.1063/5.0037092}.}}

\begin{document} 
\maketitle
\flushbottom

\section{Introduction}
In this work we investigate the canonical structure of the bosonic sector of the unique ungauged maximal supergravity theory in five dimensions that is manifestly invariant under the global action of E$_{6(6)}(\mathbb{R})$.\\

Toroidal compactification of eleven dimensional supergravity leads to symmetries described by the exceptional Lie groups E$_{n(n)}(\mathbb{R})\,\,(n=6,7,8)$ in the lower dimensional theories in $11-n$ dimensions. This fact was first discovered by Cremmer and Julia in 1979 \cite{Cremmer:1979up} and Cremmer went on to describe the five dimensional case with E$_{6(6)}(\mathbb{R})$ symmetry in detail in 1980  \cite{Cremmer}.\footnote{For more information on the exceptional Lie groups see references \cite{cartan, greenbook-lie-alg}.} The occurrence of exceptional symmetries in maximal supergravity remains one of the most remarkable features of these theories and remains to be fully understood at the quantum level.\\

We choose to analyse the E$_{6(6)}(\mathbb{R})$ invariant theory since it is one of the simplest examples of a theory with an exceptional symmetry. In five dimensions no self-dual forms exist and the symmetries are symmetries of the Lagrangian itself, not just of the equations of motion. Furthermore the group E$_{6(6)}(\mathbb{R})$ itself is easier to work with than the larger exceptional groups. Moreover we choose to analyse the bosonic sector of the theory since this greatly reduces the complexity of the analysis, while still keeping many interesting features intact.\\

\emph{E$_{6(6)}(\mathbb{R})$ exceptional field theory} --- first described in 2013 \cite{Hohm_2013First} --- is a E$_{6(6)}(\mathbb{R})$ covariant extension of eleven dimensional supergravity on a $5+27$ dimensional generalised space-time with external and internal coordinates $(x^\mu, Y^M)$ \cite{Hohm_2013First, EFTI-E6, Baguet:2015xha, Musaev:2014lna}. The internal coordinates $Y^M$ carry an index in the fundamental representation of E$_{6(6)}(\mathbb{R})$. The so called \emph{section condition} of the theory is a E$_{6(6)}(\mathbb{R})$ covariant constraint on the coordinate derivatives $\partial_M$ of the internal geometry, which requires that only a subset of the $27$ internal coordinates is physical.\\ 

Exceptional field theory has since developed into a very active field with too many applications and directions to mention here in detail. Some examples are the construction of consistent truncations and duality covariant solutions of supergravity theories (including also less than maximal supersymmetry), a unifying treatment for brane solutions and exotic branes, non-geometric backgrounds, duality covariant graviton amplitudes and many more --- we refer the reader to the following review and publications \cite{berman2020geometry,Hohm:2014qga,Bossard:2015foa}.\\

The ungauged E$_{6(6)}(\mathbb{R})$ invariant supergravity theory in five dimensions is related to the E$_{6(6)}(\mathbb{R})$ exceptional field theory by taking the trivial solution to the section condition $\partial_M=0$, thus removing all traces of the exceptional generalised geometry \cite{EFTI-E6}. This procedure is equivalent to the compactification of eleven dimensional supergravity on a six dimensional torus \cite{EFTI-E6}.\\

So far the canonical formulation of exceptional field theory has not been investigated, at least in part due to the great complexity of the exceptional generalised geometry, the complicated topological term and the many covariantisation terms involved. However a comprehensive canonical analysis of exceptional field theory should cast light on the exact role of the section condition --- in particular on its relevance to the closure of the gauge algebra. Furthermore it would explicitly demonstrate the role of the external diffeomorphisms in interconnecting the terms of the bosonic exceptional field theory Lagrangian to fix all the relative coefficients \cite{EFTI-E6}. Moreover one may see the canonical formulation as the starting point for the canonical quantisation procedure and a potential way to identify a higher dimensional Ashtekar connection \cite{PhysRevLett.57.2244,PhysRevD.36.1587,Godazgar_2014Ashtekar}. Likewise one may be interested in the canonical formulation of exceptional field theory as a tool for finding the solution to the local initial value problem for the exceptional generalised geometry as well as for finding the correct definition of the ADM energy and momentum in this setting.\\

This work is intended as a preparation and first stepping stone towards the canonical analysis of the full E$_{6(6)}(\mathbb{R})$ exceptional field theory.
We aim to make this analysis as explicit, self-contained and complete as possible, in order to provide a useful reference for future work. By analysing the E$_{6(6)}(\mathbb{R})$ supergravity theory on its own, we expect to gain insight into and greatly facilitate the canonical analysis of the full E$_{6(6)}(\mathbb{R})$ exceptional field theory --- nonetheless the E$_{6(6)}(\mathbb{R})$ supergravity theory is a very interesting theory in its own right and hence we expect there to be applications of this work that are not related to exceptional field theory.\\ 

We use the canonical formalism for this analysis because it is an algorithmic framework that can be applied to virtually all gauge theories. The canonical analysis yields much of the relevant information about a theory, such as its symmetries, gauge transformations, gauge algebra and physical degrees of freedom. Furthermore it is the starting point for the canonical quantisation of a theory and in its extended form the canonical formulation goes beyond the Lagrangian framework by making the full gauge freedom manifest \cite{dirac-lectures-QM, Henneaux-Teitelboim}.\\

The outline of this work is as follows.
In section \ref{sec:notation-conventions} we clarify the notation and conventions used in this work. In section \ref{subsec:ADM} we review the necessary principles of the Arnowitt-Deser-Misner formulation of general relativity. In section \ref{sec:supergravity} we describe the Lagrangian formulation of the E$_{6(6)}(\mathbb{R})$ invariant supergravity theory. 

We begin the construction of the canonical theory in section \ref{sec:can-analysis-I}. In section \ref{sec:can-momenta}
 we compute the canonical momenta and perform some redefinitions. We then state the canonical Hamiltonian in section \ref{sec:can-Hamiltonian}. The primary constraints are discussed in section \ref{sec:can-constraints}. The complete set of canonical constraints is constructed in section \ref{sec:can-constraints-secondary}. The total Hamiltonian is stated in section \ref{sec:tot-Hamiltonian}.
 
 We begin the canonical analysis itself in section \ref{sec:can-analysis-II}. We define the diffeomorphism weight in section \ref{sec:diff-weight} and compute all gauge transformations of all canonical coordinates in section \ref{sec:gauge-trans}. The algebra of constraints is calculated in section \ref{sec:algebra}. The extended Hamiltonian is stated in section \ref{sec:extended-Ham}. In section \ref{sec:dof} we calculate the number of physical degrees of freedom of the theory.  
 
 We conclude with a summary of the results in section \ref{sec:summary}. 
 
 For a short review of the notation and basics of canonical analysis see appendix \ref{sec:can-analysis-appendix}. In appendix \ref{sec:poisson-relations} we list Poisson bracket relations that are needed and useful in computations. In appendix \ref{appendix:other} we discuss some other useful formulae needed in the computations. In appendix \ref{appendix:coset} we discuss the treatment of the scalar coset constraints for the example of the SL(n)/SO(n) coset. 

 \newpage
   \subsection{Notation and conventions}\label{sec:notation-conventions}

   \FloatBarrier
   We work in the $1+4$ dimensional ADM decomposition (see subsection \ref{subsec:ADM}) of the five dimensional space-time and use the signature $(-++++)$.\footnote{We assume to work with a smooth, metric, Lorentzian and globally hyperbolic manifold.} Indices from the middle of the Greek alphabet, e.g. $\mu,\nu, \rho,\sigma,\tau,\dots$, denote five dimensional curved space-time indices. We use Greek indices from the beginning of the alphabet, e.g. $\alpha,\beta,\dots$, to denote five dimensional flat indices. We use Latin letters from the middle of the alphabet, e.g. $k,l,m,n,\dots$, for four dimensional curved spatial indices and Latin letters from the beginning of the alphabet, e.g. $a,b,c,d,e,\dots$, for flat spatial indices.\footnote{We exclude the letter $t$ from this list, since we use it to denote the curved time index.} The spatial flat indices are lowered and raised by the four dimensional Euclidean metric and thus their placement is irrelevant. For a simple reference of the types of space-time indices used see table \ref{tab:indices}. \\
   In addition to the letters mentioned above we will reserve the indices $0$ and $t$ to denote the flat and curved time direction, meaning $\alpha = (0,a)$ and $\mu = (t,m)$. \\
   
   Furthermore we use capitalised Latin letters, e.g. $K,L,M,N,\dots$, to denote indices of the fundamental representation of the E$_{6(6)}(\mathbb{R})$ Lie group, these indices run from $1,\dots,27$.\footnote{These are taken from the middle of the Latin alphabet, however in this work there will be no other types of capitalised Latin indices.}
\begin{table}[tbph]
\centering
\begin{tabular}{|c||c|c|}
\hline
  & flat & curved \\
\hline \hline 
$1+4$ dimensional & $\alpha,\beta,\gamma\dots$ & $\mu,\nu, \rho,\sigma,\tau,\dots$ \\\hline
4 dimensional & $a,b,c,d,e,\dots$ & $k,l,m,n,\dots$ \\
\hline
\end{tabular}
\caption{\label{tab:indices} The types of letters used for space-time indices with regard to their dimension and whether they are curved or flat.}
\end{table}

The object $\epsilon^{\mu\nu\rho\sigma\tau}$ is the totally antisymmetric Levi-Civita symbol and is independent of the metric and coordinates. We choose the convention $\epsilon^{01234} := +1$.

   \FloatBarrier

   \subsection{Arnowitt-Deser-Misner (ADM) formulation of general relativity}
   \label{subsec:ADM}
   
   In this section we state the main results concerning the ADM decomposition which we need for the canonical formulation of supergravity. For many more details on the topic see references \cite{1962ADM, MTW, Wald:1984, Nicolai:1992xx}.\\
   
   If we assume that the manifold is \textit{globally hyperbolic}, we can foliate the manifold into a set of space-like hypersurfaces and decompose the metric in the following way.\\

We write the metric in the vielbein formalism (also known as the tetrad- , local frame-  or Cartan formalism) as follows. $E_\mu{}^\alpha$ is the five dimensional frame field (or f\"unfbein) and $\eta_{\alpha\beta}$ is the Minkowski metric with signature $(-++++)$.\footnote{We use the vielbein formalism because it makes the full Lorentz symmetry manifest and this makes it possible to include the symmetry in the canonical analysis. Furthermore this formalism is necessary when coupling gravity to fermions. We will not be dealing with fermions in this work, but the vielbein formalism makes it easier to extend the work to fermions.}
\begin{equation}
G_{\mu\nu} = E_\mu{}^\alpha E_\nu{}^\beta \eta_{\alpha\beta}
\end{equation}
The following equivalent identities define the inverse vielbein.
\begin{align}
E_\mu{}^\alpha E_\alpha{}^\nu &= \delta_\mu^\nu  \label{eqn:inverse-vielbein}\\
E_\beta{}^\mu  E_\mu{}^\alpha  &= \delta_\beta^\alpha
\end{align}
By introducing the lapse function $N$, the shift vector $N^a$ and the spatial vierbein (four dimensional frame field) $e_{m}{}^a$ we can parametrise the f\"unfbein (five dimensional frame field) $E_\mu{}^\alpha$ in the following way.
\begin{equation}
E_\mu{}^\alpha =: \begin{pmatrix} 
                N & N^a \\
                0 & e_{m}{}^a\\
                \end{pmatrix}
\end{equation}
Using equation \eqref{eqn:inverse-vielbein} we find that the parametrisation of the inverse f\"unfbein is given by equation \eqref{eqn:inverse-frame-ADM}.
\begin{equation}
\label{eqn:inverse-frame-ADM}
E_\alpha{}^\mu = \begin{pmatrix} 
                N^{-1} & -N^{-1}N^m \\
                0 & e_{a}{}^m\\
                \end{pmatrix}
\end{equation}
The components of the metric that follow from these parametrisations are given in equation \eqref{eqn:ADM-metric} and following. The four dimensional metric on the spatial hypersurfaces is given by $g_{mn}$.
\begin{align}
\label{eqn:ADM-metric}
 G_{tt} &= N^a N_a - N^2 \\
 G_{tn} &= N_n := N_a e_n{}^a \\
 G_{mn} &= g_{mn} := e_m{}^a e_n{}^b \delta_{ab} 
\end{align}
When decomposing the inverse metric one should note that the components of the inverse metric are not the inverse of the components of the metric, in particular with respect to the spatial metric inverse (see equation \eqref{eqn:spatial-inv-metric}). The inverse metric components are written out as follows.
\begin{align}
\label{eqn:ADM-metric-inv}
 G^{tt} &= - N^{-2} \\
 G^{tn} &= N^{-2} N^n \\
 G^{mn} &= g^{mn} - N^{-2} N^mN^n \label{eqn:spatial-inv-metric}
\end{align}
The relation between the determinant of the five dimensional frame field $E$ and the determinant of the spatial four dimensional frame field $e$ is given by a factor of the lapse function.
\begin{equation}
E=N\cdot e
\end{equation}
The determinant of the metric $G$ can be expressed in the following form.
\begin{equation}
G = - E^2 = - N^2 e^2
\end{equation}

Other formulas relating to the geometry, that are useful in computations, can be found in the appendix \ref{appendix:other}.\\

When matter is coupled to the Einstein-Hilbert action we arrive at the ADM decomposition by inserting the decomposition of the metric or vielbein fields and then separate the matter field components by their space-time index structure. For example a one form $A_\mu$ splits into $A_t$ and $A_m$, a two form $B_{\mu\nu}$ into $B_{tn}$ and $B_{mn}$ and a scalar field remains unchanged.

  \subsection{\texorpdfstring{E$_{6(6)}(\mathbb{R})$}{E6} invariant five dimensional (super-)gravity}\label{sec:supergravity}
  
  We are interested in the bosonic sector of the maximal supergravity in five dimensions that is manifestly E$_{6(6)}(\mathbb{R})$ invariant. The bosonic field content of this theory is given by $\{G_{\mu\nu}, A^M_\mu, M_{MN} \}$ or in the ADM decomposition by $\{N, N^a, e_{m}{}^a, A^M_t, A^M_m, M_{MN} \}$ \cite{Cremmer,Freedman:2012zz}. The lapse function, the shift vector field and the spatial vielbein were already introduced in section \ref{subsec:ADM}.\\
  
  The fields $A^M_\mu$ are abelian vector gauge fields with one Lorentz index and one fundamental E$_{6(6)}(\mathbb{R})$ index. \\
  
  The fields $M_{MN}$ are Lorentz scalars and carry two fundamental E$_{6(6)}(\mathbb{R})$ indices that are symmetric. They are elements of the E$_{6(6)}(\mathbb{R})/$USp$(8)$ coset which is 42 dimensional, hence not all components of $M_{MN}$ are independent \cite{Cremmer}. 
  Due to the coset structure $M_{MN}$ transforms covariantly under the global action of E$_{6(6)}(\mathbb{R})$ and is invariant under the local action of USp$(8)$. The scalar fields can be interpreted as an E$_{6(6)}(\mathbb{R})$ metric and if one considers also the fermionic sector of the theory it is in fact necessary to also rewrite the scalar fields in terms of the 27 dimensional E$_{6(6)}(\mathbb{R})$ vielbeine (see reference \cite{Cremmer}). The E$_{6(6)}(\mathbb{R})$ metric interpretation becomes much clearer when one considers the full E$_{6(6)}(\mathbb{R})$ exceptional field theory (see references \cite{EFTI-E6, Baguet:2015xha}).
  
  We treat the E$_{6(6)}(\mathbb{R})/$USp$(8)$ coset constraints as being \emph{implicit} --- meaning that in this formalism $M_{MN}$ is a priori treated as if it is a generic symmetric matrix with 378 components ($M,N=1,...,27$) until all Poisson brackets have been evaluated.\\ This allows us to sum over all components of $M_{MN}$ without distinguishing whether they are truly independent coset parameters or not and simplifies the notation and canonical analysis significantly.
  However there are implicit coset constraints on this $M_{MN}$ that ultimately --- after evaluating all Poisson brackets --- guarantee that it is an $E_{6(6)}/\text{USp}(8)$ coset representative with 42 degrees of freedom. We want to treat the coset constraints on $M_{MN}$ implicitly because of their great complexity and because we do not need to use them explicitly in the context of this work --- one should think of them as being added to the Lagrangian with appropriate Lagrange multipliers. If we were to do this explicitly they would appear as canonical constraints and their consistency conditions would generate further canonical constraints on the canonical momenta $\Pi^{MN}(M)$. The constraints would relate various components of $M_{MN}$ and their canonical momenta $\Pi^{MN}(M)$ amongst each other --- implying among other things that $\det(M)=1$. Because they are canonical constraints we are not allowed to apply them before fully evaluating all Poisson brackets. Indeed if we treated them explicitly they would require the introduction of a Dirac bracket because the canonical coset constraints of the fields and the momenta will form a second class system of constraints. To clarify this implicit treatment of the coset constraints we demonstrate in appendix \ref{appendix:coset} the explicit and implicit formalism for the much simpler case of the coset SL(n)/SO(n) --- which has as its only coset constraint $\det(M)=1$. The scope of the implicit formalism will be sufficient for all calculations done in this work and we will find that it leads to the right gauge transformations and dynamics when comparing it to the Lagrangian formulation.\\
  
In the Lagrangian formalism the theory is given by the action of equation \eqref{eqn:action} and the Lagrangian density can be written in the form of equation \eqref{eqn:Lagrangian}. The Lagrangian density was first described in reference \cite{Cremmer}. Another way of obtaining equation \eqref{eqn:Lagrangian} is to apply the trivial ($\partial_M = 0$) solution of the section condition to the full E$_{6(6)}(\mathbb{R})$ exceptional field theory Lagrangian density, thus removing all terms related to the exceptional generalised geometry \cite{EFTI-E6}.
\begin{align}
           S = &\int d^5x \,\,  \mathcal{L}_{\text{5D}}\label{eqn:action}\\
    \mathcal{L}_{\text{5D}} =  &+ E \,\,\prescript{(5)}{} R \nonumber\\
                               &+\frac{1}{24}  E\,\,\, G^{\mu\nu}\,\,\, \partial_\mu M_{MN} \,\,\,\partial_\nu M^{MN} \nonumber \\
                               &- \frac{1}{4}  E \,\,\,M_{MN} \,\,\,F_{\mu\nu}^M \,\,\,F_{\rho\sigma}^N \,\,\,G^{\mu\rho}\,\,\, G^{\nu\sigma} \nonumber\\
                               & + \kappa \,\, \epsilon^{\mu\nu\rho\sigma\tau}\,\,d_{LMN} \,\, A^L_\mu \,\, F^M_{\nu\rho} \,\, F^N_{\sigma\tau} \label{eqn:Lagrangian}
\end{align}
The gravitational part of the theory is described by the Einstein-Hilbert term and minimal coupling to the other fields. $\prescript{(5)}{} R = \prescript{(5)}{} R^\mu{}_{\nu\mu\sigma}\,G^{\nu\sigma}$ is the Ricci scalar in five dimensions.\\

The kinetic term of the scalar fields is a non-linear sigma model of the E$_{6(6)}(\mathbb{R})/$USp$(8)$ coset --- with the coset constraints treated as explained above. The inverse scalar fields $M^{MN}$ are defined by $M_{MP} M^{PN} = \delta_M^N$.\\

The vector field kinetic term is described by a Maxwell theory type term with the additional contraction of the E$_{6(6)}(\mathbb{R})$ indices by the scalar fields. 
The abelian field strength is given by  $F_{\mu\nu}^M := 2\,\,\, \partial_{[\mu} A^M_{\nu]}$.\footnote{The Bianchi identity $\partial_{[\mu} \, F^M_{\nu\rho]}=0$ is not a canonical constraint, since it does not put any constraints on the canonical variables. The Bianchi identity simply follows from the commutativity of partial derivatives.} The abelian gauge group is U$(1)^{27}$ due to the fact that $27$ is the dimension of the fundamental representation of E$_{6(6)}(\mathbb{R})$ and we have $27$ copies of the vector field.

In addition there is a metric independent topological term of the form $A\wedge F \wedge F$ --- note that this term is also second order in derivatives. The coefficient $\kappa=+\frac{\sqrt{10}}{24}$ of the topological term (the sign is convention dependent) is needed for maximal supersymmetry in five dimensions.\footnote{One can consider other values of $\kappa$ if one is not interested in the maximally supersymmetric theory. In reference \cite{HenneauxG2} the ungauged minimal supergravity theory with the value $\kappa=\pm \frac{1}{3\sqrt{3}}$ is considered (without the scalar field terms and with a single gauge field) and yields a $G_2$ symmetry upon compactification to three dimensions.} This value of $\kappa$ also guarantees E$_{8(8)}(\mathbb{R})$ invariance upon reduction to three dimensions (cf. reference \cite{HenneauxG2} for the reduction from eleven dimensions). The precise value is not relevant for this paper and we will keep the coefficient general.

 The symbols $d_{LMN}$ and $d^{LMN}$ are fully symmetric and carry three fundamental E$_{6(6)}(\mathbb{R})$ indices. They are the (up to factors) unique invariant symbols of the fundamental representation of E$_{6(6)}(\mathbb{R})$ and we use the normalisation given by $d_{MNQ} \,\,d^{MNP} = \delta^P_Q$ \cite{Cremmer,EFTI-E6}. In this work we only need the fact that they are fully symmetric. For more details and useful identities regarding these symbols see reference \cite{EFTI-E6}.\\

All objects in the Lagrangian are in five dimensions and all objects in the following sections are written in the $1+4$ dimensional ADM decomposition of the five dimensional fields as described in section \ref{subsec:ADM}.

\section{Canonical formulation: Hamiltonian and canonical constraints}\label{sec:can-analysis-I}

In this section we construct the Hamiltonian formulation of the theory, starting from the Lagrangian formulation described in section \ref{sec:supergravity}. As a reminder or as a brief   introduction we summarise the main definitions and notations of the canonical formalism and canonical analysis in appendix \ref{sec:can-analysis-appendix}.

  \subsection{Canonical momenta}\label{sec:can-momenta}
  
We will employ the notation that a capital $\Pi(X)$ --- with appropriate indices --- signifies the canonical momentum associated to the field $X$. In many cases the specification of $X$ can be skipped since it is usually clear from the index structure of $\Pi$ alone which field is conjugate to it.\\

Calculating the canonical momenta (see equation \eqref{eq:def-can-mom} for the definition) of the lapse function, shift vector and the time component of the gauge field we immediately find the following primary constraints of the form $\Pi(X)=0$.
\begin{align}
     \Pi(N) = &  \,0  \\
     \Pi^a(N_a) = & \, 0   \\
     \Pi_M(A_t^M) = & \, 0   
\end{align} 
They vanish due to the fact that the Lagrangian density in equation \eqref{eqn:Lagrangian} does not depend on the time derivative of their conjugate fields. We refer to this type of constraint as a \emph{shift type} constraint since the gauge transformations generated by these constraints are shifts of the conjugate fields (as we will see in section \ref{sec:gauge-trans}). \\

The remaining canonical momenta do not vanish since the Lagrangian does contain time derivatives of their conjugate fields.\footnote{A dot above a function --- such as in equation \eqref{eq:mom-scalar} --- indicates a $\partial_t$ derivative operator.}
\begin{align}
     \Pi^{m}{}_a(e) = &+2\,e\, e^{bm} \big{[}\Omega_{0(ab)}-\delta_{ab}\Omega_{0cc}\big{]} \label{eq:mom-vielbein} \\
\Pi^{l}_T(A) =  &+ \frac{e}{N} g^{ln}\;M_{TN}\;\big[ F^{N}_{tn}+ N^k F^{N}_{nk}  \big] + 4\kappa\, \epsilon^{lmnr}d_{MNT}\;A^M_m\;F^N_{nr} \label{eq:mom-vector} \\
\Pi^{RS}(M) = &+\frac{1}{6} \frac{e}{N}\enspace \big{[}\dot{M}_{QP}\; M^{QR}M^{PS} +N^n\;\partial_nM^{RS}\big{]} \label{eq:mom-scalar}
\end{align}
The component of the coefficient of anholonomy $\Omega_{0bc} = N^{-1}\cdot[e_b{}^n(\partial_t-N^m\partial_m)e_{nc}-e_b{}^me_{nc}\partial_m N^n]$ in equation \eqref{eq:mom-vielbein} follows from the ADM decomposition of the definition of the coefficient of anholonomy $\Omega_{\alpha\beta\gamma} := 2\,E_{[\alpha}{}^\mu\, E_{\beta]}{}^\nu\, \partial_\mu E_{\nu\gamma}$ \,\cite{Nicolai:1992xx}.\\

It is useful to define the contractions $ \Pi_{ab}(e)$ and $\Pi(e)$ of the vielbein with the vielbein momentum.
\begin{align}
    \Pi_{ab}(e) := &+ e_{m(a} \,\Pi^m{}_{b)}(e) \\
        \Pi(e) := &+e_{m}{}^a \, \Pi^m{}_a(e)
\end{align}
To compare with the metric formulation of canonical general relativity (see e.g. references \cite{1962ADM,dewittCanGR, MTW}) one can use equation \eqref{eqn:metric-momentum-vielbein} to relate the canonical momenta of the spatial vielbein to the canonical momenta of the metric. \\
\begin{equation}\label{eqn:metric-momentum-vielbein}
    \Pi^{mn}(g)= \frac{1}{2} e^{(m}_a \Pi^{n)a}(e) 
\end{equation}
Starting from equation \eqref{eq:mom-vector} we can redefine $\Pi^{l}_T(A)$ to simplify the Hamiltonian and hence the gauge transformations. We define $P^l_L(A)$ as the canonical momentum $\Pi^{l}_T(A)$ subtracted by the topological term contribution. This way $P^l_L(A)$ takes the form expected of the theory without topological term (see equation \eqref{eq:can-mom-Yang-mills}).
\begin{align}\label{eqn:redef-P}
    P^l_L(A) := &\Pi^{l}_L(A) - 4\kappa\, \epsilon^{lmnr}d_{MNL}\;A^M_m\;F^N_{nr}  \\
            =& \frac{e}{N} g^{ln}\;M_{LN}\;\big[ F^{N}_{tn}+ N^k F^{N}_{nk}  \big]  \label{eq:can-mom-Yang-mills}
\end{align}
One has to to careful with this redefinition however since it is not a canonical transformation. This can be seen from the fact that $P^l_L(A)$ has a non-vanishing Poisson bracket with itself $\{P^l_L(A),P^k_K(A) \} \neq 0$. If we explicitly compute this Poisson bracket we find equation \eqref{PP-Bracket-E6-SUGRA}. The upper letter at the derivatives indicates the coordinate of differentiation. We refrain from using this equation in the following calculations since it is rather cumbersome --- it is nonetheless a valid identity. A more manageable approach to the calculations is to proceed order by order in the coefficient of the topological term $\kappa$.
\begin{align}
\label{PP-Bracket-E6-SUGRA}
    \{ P^l_L(A)(x) , P^k_K(A)(y) \} =
    &+ 8\kappa \,\, \epsilon^{klrs} \,\,  d_{LKM} \,\, \partial^y_r A^M_s(y) \,\,   \delta(x-y)\nonumber \\
    &+ 8\kappa \,\, \epsilon^{klrs}\,\,   d_{LKM}\,\,   A^M_r(y) \,\,  \partial^x_s \delta(x-y) \nonumber \\
    &+ 8\kappa \,\, \epsilon^{klrs} \,\,  d_{LKM} \,\, \partial_r^x A^M_s(x) \,\,    \delta(x-y) \nonumber \\
    &+ 8\kappa \,\, \epsilon^{klrs} \,\,  d_{LKM}\,\,   A^M_r(x) \,\,  \partial^x_s   \delta(x-y)  
\end{align}
Nonetheless the use of the momentum $P^l_L(A)$ greatly simplifies the Hamiltonian and it has nice transformation properties as we will see in section \ref{sec:gauge-trans}. \\

The scalar momentum $\Pi^{RS}(M)$ from equation \eqref{eq:mom-scalar} is computed using the part of the variation of the Lagrangian given by equation \eqref{eq:derivative-dot-M}. We furthermore assume equation \eqref{eqn:can-rel-Poisson-scalar-first-statement} as the definition of the fundamental Poisson bracket relation of $\Pi^{RS}(M)$ with the scalar fields. As we explained in section \ref{sec:supergravity} and appendix \ref{appendix:coset} we treat the scalar fields and their momenta as generic symmetric matrices until all the Poisson brackets have been fully evaluated and hence equation \eqref{eqn:can-rel-Poisson-scalar-first-statement} is the fundamental Poisson bracket of a symmetric matrix. Equation \eqref{eqn:can-rel-Poisson-scalar-first-statement} furthermore eliminates the need for any distinction between diagonal and off-diagonal elements of the scalar fields and momenta. We assume that sums run over the full index range.
\begin{align}
      \delta\mathcal{L}_\text{5D} = & \frac{1}{2}\, \Pi^{MN}(M)\,\,\, \delta \dot{M}_{MN} \, + \dots \label{eq:derivative-dot-M} \\
       \{M_{MN}(x), \Pi^{PQ}(M)(y) \} = & \left(\delta^P_M \, \delta^Q_N + \delta^P_N \,\delta^Q_M \right) \delta^{(4)}(x-y) \label{eqn:can-rel-Poisson-scalar-first-statement}
\end{align}
\subsection{Canonical Hamiltonian}\label{sec:can-Hamiltonian}

Having found all the canonical momenta in section \ref{sec:can-momenta}, we can now calculate the Hamiltonian density associated to equation \eqref{eqn:Lagrangian} by a Legendre transformation. To do so we calculate the ADM decomposition of all terms of the Lagrangian density --- as described in section \ref{subsec:ADM} --- and then perform the Legendre transformation with respect to all canonical momenta (see equation \eqref{eq:def-ham-density}). \\

The canonical Hamiltonian density is given in equation \eqref{eq:can-Hamiltonian}. Here we can factor out the Lagrange multipliers --- meaning the fields whose momenta are primary constraints of shift type --- thus already making the secondary constraints apparent.

\begin{align}\label{eq:can-Hamiltonian}
    \mathcal{H}_{\text{5D}}\,\, = &+ N \cdot \bigg{[} +\frac{1}{4 e} \Pi_{ab}(e)\,\,  \Pi_{ab}(e) - \frac{1}{12e} \Pi(e)^2 - e\,\,  R   \nonumber\\
                              & \hspace{1.3cm } +\frac{3}{2e} \Pi^{MN}(M) \,\, \Pi^{RS}(M)\,\,  M_{MR}\,\,  M_{NS}   - \frac{e}{24} g^{kl} \,\, \partial_k M_{MN} \,\, \partial_l M^{MN}      \nonumber \\
                              &\hspace{1.3cm }+\frac{e}{4} M_{MN}\,\,  g^{rm}\,\,  g^{sn} \,\, F_{rs}^M\,\,  F_{mn}^N       +\frac{1}{2e} g_{lm}\,\,  M^{KL} \,\, P^l_L\,\,  P^m_K \bigg{]} \nonumber \\
                          & +N^n \cdot \bigg{[}   + 2\,\, \Pi^m{}_a(e)\,\, \partial_{[n} e_{m]a} - e_{na}\,\, \partial_m \Pi^m{}_a(e)     \nonumber \\
                              &\hspace{1.5cm } +\frac{1}{2} \Pi^{MN}(M)\,\, \partial_n M_{MN}     \nonumber \\
                              & \hspace{1.5cm } + F_{nl}^M P^l_M\bigg{]}    \nonumber \\
                           & +A_t^M \cdot \bigg{[}  -\partial_l P^l_M - 3\kappa\, \epsilon^{lmnr}\,\, d_{MNP}\,\, F^N_{lm} \,\, F^P_{nr}  \bigg{]} \nonumber \\
                            &+ \dot{N} \cdot \Pi(N)  + \dot{N}_a\cdot \Pi^a(N_a) + \dot{A}_t^M \cdot\Pi_M(A_t) 
\end{align}

We can see that there are just three terms that contain the redefined momentum $P^l_M(A)$ and only one topological term ---  just like in the Lagrangian density. Thus using $P^l_M(A)$ instead of $\Pi^l_M(A)$ gives a much simpler Hamiltonian --- as can be seen from reinserting the definition of $P^l_M(A)$.\\

The terms in the last line of the Hamiltonian stem from the Legendre transformation --- since the Lagrangian does not depend on the time derivatives of the Lagrange multipliers these terms stay as they are.

It is common to work on the surface of primary constraints --- consequently removing the last line from the Hamiltonian. However we aim to keep the setting as general as possible and do not want to restrict the analysis to a subregion in phase space.

\subsection{Primary constraints}\label{sec:can-constraints}

Calculating the canonical momenta we have already seen that some of them vanish to yield primary constraints of shift type (see section \ref{sec:can-momenta}).
There are six more primary constraints --- called the \emph{Lorentz constraints} $L_{ab}$, with $L_{ab} = L_{[ab]}$. Since the Lorentz symmetry is manifest in the vielbein formalism there are constraints associated to this symmetry. The constraints $L_{ab}$ are not of shift type and are not obvious, but they do follow immediately from the momenta of the spatial vielbein (see equation \eqref{eq:mom-vielbein}) and take the explicit form of equation \eqref{eqn:Lorentz-constr} \cite{Nicolai:1992xx}.  The complete list of all $38$ primary constraints is as follows.

\begin{align}
     \Pi(N) = &\,\,0.  \\
     \Pi^a(N_a) = &\,\,0   \\
     \Pi_M(A_t^M) = &\,\,0  \\
   L_{ab}:= e_{m[a} \Pi^m{}_{b]}(e) = &\,\, 0  \label{eqn:Lorentz-constr}
\end{align} 

\subsection{Secondary constraints}\label{sec:can-constraints-secondary}

In this section we follow the procedure for finding the complete set of constraints and guaranteeing their consistency --- as outlined in appendix \ref{sec:can-analysis-appendix}.\\

In order for the primary constraints to be consistent we require that their time evolution is constant --- i.e. that their time derivative vanishes \cite{Henneaux-Teitelboim}. 
We are free to add the primary constraints --- with arbitrary phase space functions $C_0, C_1, C_2, C_3$ as parameters --- to the Hamiltonian, since we can arbitrarily extend the Hamiltonian away from the primary constraint surface in phase space. We arrive at a preliminary total Hamiltonian given by equation \eqref{eqn:prelim:total-Ham}.
\begin{equation}\label{eqn:prelim:total-Ham}
    \tilde{\mathcal{H}}_{T} := \mathcal{H}_{\text{5D}} +    C_0 \cdot \Pi(N)  + (C_1)_a\cdot \Pi^a(N_a) + (C_2)^M \cdot\Pi_M(A_t) + (C_{3})^{ab} \cdot L_{ab} 
\end{equation}
Using $ \tilde{\mathcal{H}}_{T}$ we can now test the consistency of a primary constraint $\Phi$ using equation \eqref{eqn:consistency-secondary-constr} (compare to equation \eqref{eq:secondary-constraints}).
\begin{equation}\label{eqn:consistency-secondary-constr}
    \dot{\Phi} = \{\Phi, \tilde{\mathcal{H}}_{T} \} \overset{!}{=} 0
\end{equation}
It is important to note here that the primary constraints all Poisson-commute amongst each other --- with the exception of the Lorentz constraints with themselves $\{L_{ab}, L_{cd}\} \neq 0$, since they form the Lorentz subalgebra, as we will see in section \ref{sec:algebra}.\\
 
First we consider the consistency of the primary constraints of shift type --- meaning vanishing canonical momenta. Since the primary constraints of shift type appear as Lagrange multipliers in the Hamiltonian and Poisson-commute with all other primary constraints, we find that their consistency requires that each shift type primary constraints yields one secondary constraint.
We call equation \eqref{eqn:ham-constr} the \emph{Hamilton constraint} $ H_\text{Ham} := \{\mathcal{H}_{\text{5D}} ,\, \Pi(N) \} $, equation \eqref{eqn:diff-constr} the \emph{diffeomorphism constraint} $ (H_\text{Diff})_n := \{\mathcal{H}_{\text{5D}} ,\, \Pi_a(N_a) \}\, e^a_n  $ and equation \eqref{eqn:gauss-constr} the \emph{Gau{\ss}{}\,constraint} $   (H_\text{Gau\ss})_M := \{\mathcal{H}_{\text{5D}} ,\, \Pi_M(A_t^M) \} $.\footnote{The Gau{\ss}{} constraint is named  due to the similarity to the constraint in Maxwell theory and Gau\ss's\, law. Note that we have chosen to include a vielbein in the definition of the diffeomorphism constraint in order for the constraint to have a curved index.} We will see in section \ref{sec:gauge-trans} that these names are indeed justified.
\begin{align}
    H_\text{Ham} = & +\frac{1}{4 e} \Pi_{ab}(e)\,\,  \Pi_{ab}(e) - \frac{1}{12e} \Pi(e)^2 - e\,\,  R   \nonumber \\
                              & +\frac{3}{2e} \Pi^{MN}(M) \,\, \Pi^{RS}(M)\,\,  M_{MR}\,\,  M_{NS}   - \frac{e}{24} g^{kl} \,\, \partial_k M_{MN} \,\, \partial_l M^{MN}      \nonumber \\
                              &+\frac{e}{4} M_{MN}\,\,  g^{rm}\,\,  g^{sn} \,\, F_{rs}^M\,\,  F_{mn}^N       +\frac{1}{2e} g_{lm}\,\,  M^{KL} \,\, P^l_L\,\,  P^m_K \label{eqn:ham-constr} \\
    (H_\text{Diff})_n = &+ 2\,\, \Pi^m{}_a(e)\,\, \partial_{[n} e_{m]a} - e_{na}\,\, \partial_m \Pi^m{}_a(e)     \nonumber \\
                              & +\frac{1}{2} \Pi^{MN}(M)\,\, \partial_n M_{MN}     \nonumber \\
                               &  + F_{nl}^M P^l_M\label{eqn:diff-constr} \\
    (H_\text{Gau\ss})_M = &  -\partial_l P^l_M - 3\kappa\, \epsilon^{lmnr}\,\, d_{MNP}\,\, F^N_{lm} \,\, F^P_{nr} \label{eqn:gauss-constr}
\end{align}

The secondary constraints restrict the canonical coordinates dynamically since we have made use of the equations of motion to find them. The secondary constraints are equivalent to the time components of the Lagrangian equations of motion. The Hamilton constraint is the time-time-component of the Einstein equation, the diffeomorphism constraints are the time-spatial-components of the Einstein equation. Similarly the Gau{\ss}{} constraints are the time-components of the analog of the Maxwell equation. \\

It is worth noticing that the Gau{\ss}{} constraints $H_\text{Gau\ss}$ are independent of the metric and scalar degrees of freedom.\\

The gravitational part of the diffeomorphism constraint $H_\text{Diff}$ can be rewritten using equation \eqref{eq:diffeo-constr-Lorentz} to reveal a term containing the Lorentz constraint with a field dependent coefficient given by the spin connection $\omega_{n}{}^{ab} = e^{ak} \nabla_n e_k{}^b $. The covariant derivative $\nabla_n$ contains the Levi-Civita connection and $D_m$ contains the spin connection. 
\begin{equation} \label{eq:diffeo-constr-Lorentz}
    + 2\,\, \Pi^m{}_a(e)\,\, \partial_{[n} e_{m]a} - e_{na}\,\, \partial_m \Pi^m{}_a(e)  = -e_n{}^a \,D_m\Pi^m{}_a(e) + \, \omega_{n}{}^{ab}\,L_{ab}
\end{equation}
One can redefine the diffeomorphism constraint as in equation \eqref{eq:diffeo-constr-redef}, which makes the new diffeomorphism constraint $\tilde{H}_\text{Diff}$ and the Lorentz constraint Poisson-commute. However we will see that the constraint of equation \eqref{eqn:diff-constr} gives the nicer and expected gauge transformations (see equation \eqref{eqn:diffeo-on-vielbein}). We will thus continue to work with the diffeomorphism constraint of equation \eqref{eqn:diff-constr}. The redefinition would simply represent a different choice of basis of the constraint algebra.
\begin{equation} \label{eq:diffeo-constr-redef}
     (\tilde{H}_\text{Diff})_n :=  (H_\text{Diff})_n - \, \omega_{n}{}^{ab}\,L_{ab} 
\end{equation}

The only primary constraints not of shift type are the Lorentz constraints. If we insert the Lorentz constraints in equation \eqref{eqn:consistency-secondary-constr} we find that no new constraints are being generated. The consistency requirement does however restrict the coefficient $(C_{3})^{ab}$ of the Lorentz constraint term $(C_{3})^{ab} \, L_{ab}$ in the total Hamiltonian. The coefficient has to be $(C_{3})^{ab} = -N^n\, \omega_{n}{}^{ab}$ for the constraints to be consistent. This is precisely the term found in  equation \eqref{eq:diffeo-constr-redef}. This means that the total Hamiltonian does not contain an intrinsic Lorentz constraint term. We will see in section \ref{sec:extended-Ham} that there is a Lorentz constraint term in the extended Hamiltonian.\\

To verify the consistency of the secondary constraints we also take equation \eqref{eqn:consistency-secondary-constr} and now insert the secondary constraints as $\Phi$. We find no tertiary constraints implying that the set of constraints we have found so far is complete and consistent. There is no constraint associated to the exceptional E$_{6(6)}(\mathbb{R})$ symmetry of the theory since it is a global symmetry of the theory and not a gauge symmetry.\\

It is beneficial to make use of the concept of \emph{integrated} or \emph{smeared constraints} in order to avoid expressions that contain derivatives of the delta distribution. We define the smeared constraints by contracting all indices of the constraints with a tensor of test functions with the same symmetries and then integrating this expression over the entire spatial hypersurface. We denote the integrated constraint by adjoining brackets with the name of the test function to the constraint.\\ 

 If we take the diffeomorphism constraint as an example, the smeared constraint is defined by equation \eqref{eqn:smeared-constraint}, where $\lambda^n(x)$ is a vector of test functions on the spatial hypersurface. 
\begin{equation}\label{eqn:smeared-constraint}
     H_\text{Diff}[\lambda] \,\,:=\,\,  \int\,\, (H_\text{Diff})_n(x)\,\, \lambda^n(x)\,\, d^4x
\end{equation}
Note that no symmetry factors are inserted. The smeared version of the Lorentz constraints is given by equation \eqref{eq:smeared-Lorentz}, where $\gamma^{ab}=\gamma^{[ab]}$.
\begin{equation}\label{eq:smeared-Lorentz}
    L[\gamma] \,\,:=\,\,  \int\,\,L_{ab}(x)\,\, \gamma^{ab}(x)\,\, d^4x
\end{equation}

From equation \eqref{eq:can-Hamiltonian} we can now moreover see that the Hamiltonian is weakly vanishing $\mathcal{H}_{\text{5D}} \approx 0$, demonstrating that the theory is generally covariant \cite{Henneaux-Teitelboim}.

\subsection{Total Hamiltonian}\label{sec:tot-Hamiltonian}

The total Hamiltonian is given by equation \eqref{SUGRA-eq:total-Hamiltonian}. To arrive at the total Hamiltonian we start with the canonical Hamiltonian of equation \eqref{eq:can-Hamiltonian} and add all primary constraints with coefficients that satisfy the consistency conditions derived in section \ref{sec:can-constraints-secondary}.
\begin{align}\label{SUGRA-eq:total-Hamiltonian}
    \mathcal{H}_{\text{T}}\,\, =  &+N \cdot  H_\text{Ham} +N^n \cdot  (H_\text{Diff})_n+A_t^M \cdot  (H_\text{Gau\ss})_M \nonumber\\
&+ C_0 \cdot \Pi(N)  + (C_1)_a\cdot \Pi^a(N_a) + (C_2)^M \cdot\Pi_M(A_t) \nonumber\\
& -N^n\, \omega_{n}{}^{ab} \, L_{ab}
\end{align}

The coefficients $C_0, (C_1)_a$ and $(C_2)^M$ are arbitrary phase space functions. The coefficient of the Lorentz constraint term is restricted to this particular form containing the spin connection as explained in the previous subsection. \\

The time evolution of a phase space function $F$, generated by the total Hamiltonian, via equation \eqref{eqn:time-evo-total-Ham} is equivalent to the original Lagrangian time evolution \cite{Henneaux-Teitelboim}. Note that the equality in equation \eqref{eqn:time-evo-total-Ham} is \emph{weak} --- this means that after evaluating the Poisson bracket the equation holds only on the phase space surface where the constraints are satisfied (see appendix \ref{sec:can-analysis-appendix}).
\begin{equation}\label{eqn:time-evo-total-Ham}
    \dot{F} \approx \{F, \mathcal{H}_T \}
\end{equation}
In section \ref{sec:extended-Ham} we construct the extended Hamiltonian and find that the exact coefficients in the total Hamiltonian are irrelevant to the most general time evolution since they are all replaced by arbitrary coefficient functions in the extended Hamiltonian.\\

In practice it is often easiest to determine the time evolution by first computing all the gauge transformations. We compute all gauge transformations of all fields and momenta in section \ref{sec:gauge-trans}. 
   
\section{Canonical analysis: Gauge transformations and gauge algebra}\label{sec:can-analysis-II}

 In this section we analyse the constraints we found in section \ref{sec:can-analysis-I}. We calculate the gauge transformations generated by the constraints, compute the algebra that is formed by the constraints under the Poisson bracket, determine the number of physical degrees of freedom and discuss the extended Hamiltonian.

  \subsection{Diffeomorphism weight and the Lie derivative}\label{sec:diff-weight}
  
  We briefly define the diffeomorphism weight in this section, as the concept is required for the following sections.\\

The Lie derivative of a tensor $T$ with parameter $\xi$ is denoted by $\mathcal{L}_\xi T$. The \emph{diffeomorphism weight} $\Lambda(T)$ is defined as the coefficient of the weight term in the Lie derivative. For example if $T$ is a vector, we write equation \eqref{eq:lie-deriv-vector} for the components of the Lie derivative.

\begin{equation}
\label{eq:lie-deriv-vector}
    (\mathcal{L}_\xi T)^\nu = \underbrace{\xi^\mu \,\partial_\mu T^\nu}_\text{transport term}  - \underbrace{\partial_\mu\xi^\nu \, T^\mu }_\text{rotation term} +\underbrace{\Lambda\cdot\partial_\mu \xi^\mu}_\text{weight term}
\end{equation}

Table \ref{diff-weights-e6-sugra} lists the diffeomorphism weights of all the relevant fields and momenta. The vielbein determinant is a tensor density and since the Lagrangian includes a vielbein determinant the canonical momenta have diffeomorphism weight one too.

\begin{table}[!ht]
\centering 
\begin{tabular}{|c|c|}
\hline
Object & Weight $\Lambda$ \\\hline \hline
$e_m{}^a$, $A^{M}_\mu$, $M_{MN}$ & 0\\\hline
$\Pi^{m}{}_a(e)$ , $P_M^m(A)$ , $\Pi^{RS}(M)$ & 1 \\\hline
$e$ & 1 \\\hline
\end{tabular}
\caption{\label{diff-weights-e6-sugra}
The diffeomorphism weights of the most important objects.}
\end{table}
\FloatBarrier

When Lie derivatives are used in the following sections they always include the weight term appropriate for the object that the derivative is applied to.

\subsection{Gauge transformations}\label{sec:gauge-trans}
  
In this section we compute the explicit form of all infinitesimal gauge transformations. To do so we calculate the Poisson brackets of all the constraints $\Phi[\lambda]$ with all the canonical coordinates $X$ via $\delta  X = \{X, \Phi[\lambda]  \}$ (see equation \eqref{eq:gauge-transf-F}). We use the notion of smeared constraints in the following sections as explained in section \ref{sec:can-constraints-secondary}. We only state the non-vanishing gauge transformations. \\

To compute Poisson brackets of the canonical coordinates we define the fundamental Poisson brackets as follows.\footnote{Due to the implicit treatment of the coset constraints --- as explained in section \ref{sec:supergravity} and appendix \ref{appendix:coset} --- equation \eqref{eqn:can-rel-Poisson-scalar} is the fundamental Poisson bracket relation for a generic symmetric scalar matrix and thus there is no coset projector term.}
\begin{align}
      \{ N(x) , \Pi(N)(y)\}  = & \,\delta^{(4)}(x-y) \\
      \{ N^n(x) , \Pi_m(N^k)(y)\}  = & \, \delta^n_m \delta^{(4)}(x-y) \\
      \{ e_n{}^a(x) , \Pi^m{}_b(e)(y)\}  = & \, \delta_n^m \delta^a_b \delta^{(4)}(x-y) \\
      \{ A^M_t(x), \Pi_N(A^K_t)(y)\}  = & \, \delta^M_N \delta^{(4)}(x-y) \\
      \{ A^M_m(x), \Pi^n_N(A^K_k)(y)\}  = & \{ A^M_m(x), P^n_N(y)\}= \, \delta^M_N \delta^n_m \delta^{(4)}(x-y) \\
    \{M_{MN}(x), \Pi^{PQ}(M)(y) \} = & \left(\delta^P_M \, \delta^Q_N + \delta^P_N \,\delta^Q_M \right) \delta^{(4)}(x-y) \label{eqn:can-rel-Poisson-scalar}
\end{align}

Note that the redefinition of the momentum of $A^M_m$ from equation \eqref{eqn:redef-P} does not affect the fundamental Poisson bracket, since the term that is subtracted in the redefinition only depends on the field $A^M_m$.

In appendix \ref{sec:poisson-relations} we list many other useful Poisson bracket identities that are needed in the computation of the gauge transformations.

To keep the expressions as simple as possible we omit the notation of the coordinate dependence in the following. It is understood that the gauge transformation only depends on the coordinate of the field that the constraint acts upon.\\

We begin with the primary constraints of shift type. The only gauge transformations that one can generate using these constraints are shift transformations on the fields canonically conjugate to the vanishing momenta. 
\begin{align}
   \{ N , \Pi(N)[\lambda_1] \} = & \lambda_1 \\
    \{ N_a, \Pi(N_b)[\lambda_2] \}= &(\lambda_2)_a \\
     \{ A_t^N, \Pi(A_t^M)[\lambda_3] \} = &(\lambda_3)^N
\end{align} 
The Lorentz constraints generate Lorentz transformations on the spatial vielbein and its canonical momentum. The vielbein determinant is Lorentz invariant --- as are all quantities that can be expressed solely through the metric tensor. The transformations take the form of a rotation of the flat index by the smearing tensor.
\begin{align}
      \{   e_n{}^a, L[\gamma] \}= &+ e_{nb}\,\, \gamma^{ba} \\
       \{    \Pi^n{}_a(e),  L[\gamma] \}= & +\Pi^n{}_c(e)   \,\, \gamma^{cb}\,\,\delta_{ba}
\end{align}
The Hamilton constraint generates time evolution, which we can interpret as a gauge transformation \cite{Henneaux-Teitelboim}. The time evolution generated by just the Hamilton constraint --- unlike the total Hamiltonian --- does not capture any other gauge freedom. However without the Hamilton constraint time evolution is not possible at all.
The time evolution of the fields generated by the Hamilton constraint is essentially given by the canonically conjugate momenta. The time evolution of the canonical momenta is more complicated and captures most of the dynamics.
Since the vielbein --- or equivalently the spatial metric ---  contracts all terms in the Lagrangian --- with the exception of the topological term --- its conjugate canonical momentum $\Pi^n{}_a(e)$ has a particularly complicated time evolution (see equation \eqref{eqn:pi-of-e-time-evolution}).\footnote{It is useful to first compute the Poisson brackets of equation \eqref{sugra:eqn:poisson-Ricci-Pie} and the following equations.} In the third line of equation \eqref{eqn:pi-of-e-time-evolution} we see the spatial Einstein equation in the vielbein form.

The time evolution of $\Pi^n{}_a(e)$ can be compared to the time evolution of the canonical momentum of the metric in pure general relativity (see references \cite{1962ADM, MTW}) using equation \eqref{eqn:metric-momentum-vielbein}.  

\begin{align}
   \{  e_{na} ,H_\text{Ham}[\phi] \} = & +\frac{\phi}{2e} \,\, g_{mn} \,\,\Pi^m{}_a(e) - \frac{\phi}{6e}\,\,\Pi(e)\,\,e_{na}  \\
    \{ \Pi^n{}_a(e), H_\text{Ham}[\phi]  \} = & + \frac{\phi}{4e}\,\,\Pi_{bc}(e)\,\,\Pi_{bc}(e)\,\, e_a{}^n -\frac{\phi}{2e}\Pi^k{}_b(e)\,\,\Pi^n{}_b(e)\,\,e_{ka} \label{eqn:pi-of-e-time-evolution}\\
   & -\frac{\phi}{12e}\,\,\Pi^2(e)\,\,e_a{}^n  + \frac{\phi}{6e}\,\,\Pi(e)\,\,\Pi^n{}_a(e) \nonumber\\
    &-2\phi e \left( R^{nk} \,\,e_{ka} - \frac{1}{2}R\,\,e_a{}^n\right) \nonumber\\
    &  +2e \left( \nabla_a \nabla^n \phi - \nabla^k \nabla_k \phi  \,\, e_a{}^n \right)   \nonumber \\
     & +\frac{3\phi}{2e} \Pi^{MN}(M)\,\,\Pi^{RS}(M)\,\,M_{MR}\,\,M_{NS}\,\,e_a{}^n  \nonumber\\
      &  + \frac{\phi e}{24}\partial_k M_{MN}\,\,\partial_l M^{MN} \,\,g^{kl}\,\,e_a{}^n -   \frac{\phi e}{12} \partial_k M_{MN}\,\,\partial_l M^{MN}\,\,g^{ln}\,\,e_a{}^k  \nonumber\\
       &- \frac{\phi e}{4} M_{MN}\,\,F^M_{rs}\,\,F^N_{km}\,\,g^{rk}\,\,g^{sm}\,\,e_a{}^n +    \phi e \,\,M_{MN}\,\,F^M_{rs}\,\,F^N_{km}\,\,g^{rk}\,\,g^{mn}\,\,e_a{}^s  \nonumber\\
        &+ \frac{\phi}{2e} M^{KL}\,\,P^l_L(A)\,\,P^k_K(A)\,\,g_{lk}\,\,e_a{}^n -     \frac{\phi}{e}M^{KL}\,\,P^n_K(A)\,\,P^l_L(A)\,\,e_{la}  \nonumber
\end{align}
\begin{align}
         \{ M_{MN},  H_\text{Ham}[\phi]  \} = & +\frac{6}{e} \phi\,\, \Pi^{QP}(x)\,\, M_{MQ}\,\,M_{NP}  \\
\{\Pi^{MN}(M) , H_\text{Ham}[\phi] \} = & - \partial_l\left( \frac{\phi e}{6}\,\,g^{kl}\,\,\partial_kM^{MN}\right) \nonumber \\
   &  - \frac{\phi e}{6} \,\, g^{kl} \,\,\partial_kM_{KL}\,\,\partial_lM^{KM}\,\,M^{LN}   \nonumber \\
   & - \frac{6 \phi }{e} \,\,  \Pi^{PM}(M)\,\, \Pi^{NR}(M)\,\,M_{PR}   \nonumber \\
   & - \frac{\phi e}{2} \,\, g^{rm}\,\, g^{sn}\,\,F^{M}_{rs}\,\,F^{N}_{mn}  \nonumber \\
   &  + \frac{\phi}{e}  g_{lm} \,\, P^l_L \,\, P^m_K \,\, M^{KM}\,\, M^{LN} \\
      \{ A^N_n, H_\text{Ham}[\phi]  \} = & + \frac{\phi}{e} \,\, g_{nl} \,\, M^{NL} \,\, P^l_L \\
       \{  P^l_S , H_\text{Ham}[\phi]\} = &+ \partial_m \left( e\,\phi\,\, M_{NS} \,\,g^{rm} \,\,g^{ls} \,\,F^N_{rs}   \right) \\
   &- \frac{12\kappa\phi}{e} \,\,g_{mk} \,\, M^{KL}\,\, d_{SLM}\,\,\epsilon^{lmrs}\,\,F^M_{rs} \,\,P^k_K \nonumber
\end{align}

The Gau{\ss}{} constraint generates abelian U$(1)^{27}$  gauge transformations on the one form gauge field $A^N_n$. The conjugate momentum $P^n_N$ is invariant under these transformations --- like it is in the free theory --- which is a nice property of the redefinition from equation \eqref{eqn:redef-P}. Since the constraint is independent of the metric and scalar fields they do not transform.
\begin{align}
   \{ A^N_n, H_\text{Gau\ss}[\zeta] \} =   & +\partial_n \zeta^N   \\
     \{ P^n_N, H_\text{Gau\ss}[\zeta] \} = & \,0
\end{align}
The diffeomorphism constraint generates diffeomorphisms on the spatial hypersurface via the Lie derivative (including the appropriate weight terms). If we were to use the redefined diffeomorphism constraint from equation \eqref{eq:diffeo-constr-redef} we would see additional terms in the transformation of the vielbein and its conjugate momentum.
\begin{align}
   \{  e_n{}^a,H_\text{Diff}[\xi] \}= &+ \mathcal{L}_\xi e_n{}^a  \label{eqn:diffeo-on-vielbein}\\
   \{  \Pi^n{}_a(e), H_\text{Diff}[\xi] \}= &+ \mathcal{L}_\xi  \Pi^n{}_a(e)\\
    \{M_{MN}, H_\text{Diff}[\xi] \}= & +\mathcal{L}_\xi M_{MN}\\
     \{\Pi^{MN}(M),  H_\text{Diff}[\xi] \}= & + \mathcal{L}_\xi  \Pi^{MN}(M) \\
      \{ A^N_n, H_\text{Diff}[\xi] \}= &-\xi^m\,\, F^N_{nm}  \\
     = & + \mathcal{L}_{\xi}A^N_n +\delta_{(\xi^m A_m)}A^N_n \nonumber\\
      \{P^l_S, H_\text{Diff}[\xi]  \}= & + \mathcal{L}_\xi P^l_S + \xi^l\, (H_\text{Gau\ss})_S \\
   \approx & + \mathcal{L}_\xi P^l_S \nonumber 
\end{align}
Due to the parametrisation of the Lagrangian the transformation of the one form gauge field is a Lie derivative only up to a U$(1)^{27}$ gauge transformation generated by the Gau{\ss}{} constraint $H_\text{Gau\ss}$. The notation is taken to mean $\delta_{(\xi^m A_m)}A^N_n = \{A^N_n , H_\text{Gau\ss}[\xi^m A^M_m] \} = + \partial_n (\xi^m A_m^N)$.
In the case of its conjugate canonical momentum $P^l_S$ the transformation is a Lie derivative up to the Gau{\ss}{} constraint $H_\text{Gau\ss}$ and thus weakly equal to the Lie derivative.

The Schouten identity from appendix \ref{appendix:other} has to be used repeatedly to compute the transformation of the one form field and its momentum.

  \subsection{Algebra of constraints}\label{sec:algebra}
  
  The algebra that is spanned by the canonical constraints under the Poisson bracket is equivalent to the algebra of gauge transformations. We can interpret a Poisson bracket of two canonical constraints as the commutator of two infinitesimal gauge transformations.\\

  In general it is easiest to explicitly write out the simpler looking constraint and then make use of the gauge transformations from section \ref{sec:gauge-trans} to compute the algebra. For relations that involve the diffeomorphism constraint it is easiest to make use of the fact that all fields --- except the one form gauge field and its conjugate momentum --- transform as Lie derivatives under the diffeomorphism constraint.  
  
  The primary constraints of shift type Poisson-commute with all the other constraints and are therefore not listed below. The full constraint algebra can be written as follows.

\begin{align}
\{ H_\text{Ham}[ \theta] , H_\text{Ham}[ \tau]\}        &=   H_\text{Diff}[(\theta\,\nabla_m\tau-\tau\,\nabla_m\theta) \, g^{mn}]  \nonumber\\ 
                                                           &\quad- L\left[ (\theta\,\nabla_m\tau-\tau\,\nabla_m\theta) \, g^{mn}\, \omega_{nab} \right]    \label{5DSUGRA-eq:alg-ham-ham}\\
\{ H_\text{Diff}[\lambda],H_\text{Ham}[ \theta] \}     &=  H_\text{Ham}[\mathcal{L}_\lambda\theta] + H_\text{Gau\ss}\left[\frac{\theta}{e}\lambda^p\,\,g_{pk}\,\,P^k_L\,\,M^{LM}\right]  \label{5DSUGRA-ALG-Diff-Ham}\\  %c.f. A.H. Diaz
\{  H_\text{Ham}[ \theta] ,  H_\text{Gau\ss}[\xi]  \}   &=  0  \\ 
\{ H_\text{Diff}[\lambda] ,  H_\text{Diff}[\kappa] \}   &= H_\text{Diff}\left[\left[\lambda,\kappa\right]^n\right] = H_\text{Diff}\left[\mathcal{L}_\lambda\kappa^n\right]   \label{eqn:alg-diff-diff}\\  
\{H_\text{Diff}[\lambda] , H_\text{Gau\ss}[\xi] \}      &= 0   \\
\{ H_\text{Gau\ss}[\xi] , H_\text{Gau\ss}[\zeta] \}     &= 0  \label{5DSUGRA-ALG-GG}\\
\{L[\gamma] ,L[\kappa] \}      &=   L[-2 \gamma^{c[a}\,\,\kappa^{b]c}]  \label{eqn:alg-LL}\\ 
\{ H_\text{Ham}[\phi] ,  L[\gamma]  \}   &=  0   \\ 
\{ H_\text{Diff}[\lambda] ,  L[\gamma]  \}   &=     L[\mathcal{L}_\lambda(\gamma^{ab})] \label{eqn:alg-diff-L}\\  
\{ H_\text{Gau\ss}[\xi] , L[\gamma]  \}   &=   0  
\end{align}

From equation \eqref{5DSUGRA-eq:alg-ham-ham} we see that two different orderings of time evolutions (as generated by the Hamilton constraint) can only differ by a diffeomorphism and a Lorentz transformation. This means that the time evolution with the Hamilton constraint is unique up to these gauge transformations.

Due to the use of the vielbein formalism and the choice of diffeomorphism constraint (see the discussion in section \ref{sec:can-constraints-secondary}) the equation \eqref{5DSUGRA-eq:alg-ham-ham} contains a Lorentz constraint term whose smearing function depends on the spin connection. 
Nonetheless Lorentz invariance is preserved since both the spin connection and the Lorentz constraint itself transform under a Lorentz transformation to cancel out the transformation of the diffeomorphism constraint term.\footnote{
From equation \eqref{eq:diffeo-constr-redef} one can moreover see that this additional term disappears when using the redefined constraint. However the redefinition has many other consequences which are not nice, including the introduction of additional terms in the algebra and in the gauge transformations as we have already mentioned.}\\

In equation \eqref{5DSUGRA-ALG-Diff-Ham} we see that the Poisson bracket of a diffeomorphism and time evolution is the time evolution with a smearing function given by the Lie derivative of the parameter. There is furthermore a Gau{\ss}{} constraint term, which is due to the transformation property of the one form gauge field and its conjugate momentum under the diffeomorphism constraint.\\

The equations \eqref{5DSUGRA-eq:alg-ham-ham} and \eqref{5DSUGRA-ALG-Diff-Ham} moreover show that the algebra is actually an \emph{open-} or \emph{pseudo-algebra}, since the smearing functions on the right hand side depend on canonical coordinates --- see references \cite{Henneaux-Teitelboim, Freedman:2012zz} for more information. Nonetheless the term algebra is commonly used in such cases. 

Note that even in pure general relativity the smearing function of the diffeomorphism constraint in equation \eqref{5DSUGRA-eq:alg-ham-ham} contains the inverse metric. The constraint algebra of pure general relativity has been discussed in the references \cite{dewittCanGR, HOJMAN197688, Kiefer:2004gr}. In addition the canonical formulation and constraint algebra of 11 dimensional supergravity has been discussed in references \cite{PhysRevD.33.2801, PhysRevD.33.2809}.\\

The equations \eqref{eqn:alg-diff-diff}, \eqref{5DSUGRA-ALG-GG} and \eqref{eqn:alg-LL} show that diffeomorphisms, U$(1)^{27}$ gauge transformations and Lorentz transformations each form a proper subalgebra of their own. Equation \eqref{5DSUGRA-ALG-GG} thus also proves that the gauge field is indeed abelian. \\

Equation \eqref{eqn:alg-diff-L} tells us that the Poisson bracket of the diffeomorphism constraint and the Lorentz constraint is given by a Lorentz transformation with the Lie derivative of the parameter.\footnote{Using the redefinition of equation \eqref{eq:diffeo-constr-redef} one can make this term vanish, however with the same unintended consequences as mentioned before.}\\

We now know the full algebra of constraints and see that it does indeed close under the Poisson bracket. This fact implies that all canonical constraints Poisson-commute weakly --- i.e. up to constraint terms the Poisson brackets of any two constraints vanish --- or equivalently that all constraints are first class.\footnote{This is no longer true once one considers the full theory including fermions.}

 \subsection{Extended Hamiltonian}\label{sec:extended-Ham}

The extended Hamiltonian is constructed from the total Hamiltonian (equation \eqref{SUGRA-eq:total-Hamiltonian}) by adding all first class constraints with arbitrary coefficients. Since all the constraints are first class the extended Hamiltonian is the linear sum over all constraints with arbitrary coefficient functions (see equation \eqref{SUGRA-eq:extended-Hamiltonian}).
The difference between the total and the extended Hamiltonian is in this case that the parameters of the secondary constraints and the Lorentz constraints become completely arbitrary, hence allowing for more general gauge transformations than the Lagrangian time evolution. 
\begin{align}\label{SUGRA-eq:extended-Hamiltonian}
    \mathcal{H}_{\text{E}}\,\, =  &+C_\text{Ham} \cdot  H_\text{Ham} +(C_\text{Diff})^n \cdot  (H_\text{Diff})_n+(C_\text{Gau\ss})^M \cdot  (H_\text{Gau\ss})_M \nonumber\\
&+ C_0 \cdot \Pi(N)  + (C_1)_a\cdot \Pi^a(N_a) + (C_2)^M \cdot\Pi_M(A_t) \nonumber\\
&+ (C_3)^{ab} \, L_{ab}
\end{align}
The most general time evolution of a phase space function $F$ is then given by equation \eqref{eq:time-evolution-extended-hamiltonian-main}. Since we already know all the gauge transformations we can use this to efficiently compute the time evolution of any function using equation \eqref{SUGRA-eq:extended-Hamiltonian} and the linearity of the Poisson bracket.
  \begin{equation}\label{eq:time-evolution-extended-hamiltonian-main}
      \dot{F}(q,p) \approx \{F, \mathcal{H}_E \}
  \end{equation}
 The time evolution described by the extended Hamiltonian and the canonical Hamiltonian are equivalent for observables, since they are by definition gauge invariant.

 \subsection{Counting the degrees of freedom}\label{sec:dof}
  
  With the full set of constraints as well as the constraint algebra known we can re-derive the number of physical degrees of freedom of the theory. In table \ref{constraints-counting-5D-SUSY} we list all the field degrees of freedom and the primary and secondary constraints. 
  
\begin{table}[!ht]\centering
\label{constraints-counting-5D-SUSY}
\begin{tabular}{||c|c||c|c||c|c||}
\hline
\textbf{Fields} & \textbf{\#} & \textbf{Primary constraints} & \textbf{\#}  & \textbf{Secondary constraints} & \textbf{\#}  \\ \hline\hline
 $N$ & $1$ & $\Pi(N)$ & 1  & Hamilton constraint & 1  \\ \hline
 $N_a$& $4$ & $ \Pi(N_a)$ & 4 & Diffeomorphism constraints  &  4 \\ \hline
 $e_{ma}$& $16$ & Lorentz constraints & 6    & - & 0 \\ \hline 
 $M_{(MN)}$& $42$ & - & 0 & - &  0\\ \hline
 $A^{T}_t$& $27$ & $ \Pi(A^{T}_t)$ & 27  & Gau{\ss}{} constraints & 27  \\ \hline
 $A^T_l$& $108$ & - & 0 &-  & 0 \\ \hline\hline
 \textbf{Total:} & \textbf{198} & \textbf{Total:} & \textbf{38}   &  \textbf{Total:} & \textbf{32} \\ \hline
\end{tabular}
\caption{A counting of the number of fields and the number of primary and secondary canonical constraints in the bosonic sector of \texorpdfstring{E$_{6(6)}(\mathbb{R})$}{E6} invariant five dimensional supergravity. The distinction between primary and secondary constraints is irrelevant to the number of physical degrees of freedom, but illustrates where the constraints came form.}
\end{table}

We count a total of $198$ field variables --- or $396=2\cdot198$ canonical coordinates in phase space. To arrive at this result we have applied the implicit coset constraints to the scalar fields --- as outlined in appendix \ref{appendix:coset} --- to reduce the number of scalar fields to the 42 independent variables we need to describe the E$_{6(6)}(\mathbb{R})$/USp(8) coset. We also count a total of $70=38+32$ canonical constraints, all of which are first class constraints and hence have to be counted twice \cite{Henneaux-Teitelboim}. Thus the theory has $396-2\cdot70=256$ physical dimensions in phase space or equivalently $128$ physical (bosonic) degrees of freedom. This is in agreement with the well-known result that --- at each point in space --- maximal supergravity has 128 bosonic degrees of freedom \cite{Strathdee:1986jr, Cremmer, Cremmer:1978km, Freedman:2012zz}.

\section{Summary and outlook}\label{sec:summary}

Starting from the Lagrangian formulation of the bosonic sector of the E$_{6(6)}(\mathbb{R})$ invariant supergravity theory in five dimensions we have constructed the Hamiltonian formulation of that theory. We then constructed the complete set of $70$ canonical constraints. We calculated the gauge transformations generated by the canonical constraints and found that they generate time evolution, diffeomorphisms, Lorentz transformations, U$(1)^{27}$ gauge transformations and shift transformations. As is to be expected the E$_{6(6)}(\mathbb{R})$ symmetry is not generated by canonical constraints since it is a global symmetry. We found that the algebra of gauge transformations closes and that all constraints of the bosonic theory are first class. Hence the extended Hamiltonian --- describing the most general time evolution of the theory, where the full gauge freedom is manifest --- was constructed by summing over the complete set of constraints with fully arbitrary coefficient functions. We finally confirmed the well-known result that the number of physical degrees of freedom of the theory is $128$.\\

This work is intended to lay the foundation for the canonical analysis of the full E$_{6(6)}(\mathbb{R})$ exceptional field theory. Due to the complexity of the exceptional generalised geometry, the complicated topological term --- so far also lacking an explicit non-integral expression --- and the many covariantisation terms involved, the complete treatment of the exceptional generalised geometry is beyond the scope of this preparatory work. As a result of this preparation we can now focus on the more difficult tasks of understanding the internal exceptional geometry and working out its canonical structure. As was explained in the introduction there are many open questions one can ask about canonical exceptional field theory and the exceptional generalised geometry. Perhaps one of the most intriguing questions concerns the exact role of the section condition --- notably its relevance to the closure of the gauge algebra. To cast light on this and other topics we plan to proceed with the canonical analysis of the full E$_{6(6)}(\mathbb{R})$ exceptional field theory in a future publication.

\appendix

\section{Basics of canonical analysis}\label{sec:can-analysis-appendix}

In this section we briefly review the basics of how to analyse a theory using the canonical formalism.\footnote{Canonical analysis is a synonym for Hamiltonian analysis.} A full introduction to this subject is far beyond the scope of this work and we omit many details in this section. For an in depth treatment please see references \cite{dirac-lectures-QM, Henneaux-Teitelboim}.\footnote{This section is mainly based on the first chapters of reference \cite{Henneaux-Teitelboim}.}\\

The starting point of the canonical analysis is most often the Lagrangian formulation of a theory. Starting from an action functional $S$, given by the $d$-dimensional space-time integral over some Lagrangian density $\mathcal{L}(q,\dot{q})$, which in turn depends only on some fields $q^n(x)$ and their time derivatives $\dot{q}^n(x)$ (see equation \eqref{eq:S-Lagrangian}). The space-time coordinates are $x^\mu$ with indices $\mu=0,\dots,d-1$. Note that in this section the Latin indices are used to label fields and constraints of the theory.\\

The canonical formalism aims to treat the fields and their conjugate canonical momenta on an equal footing. 
We introduce the canonical momenta $p_n$ as defined by equation \eqref{eq:def-can-mom}. In the case of a field theory the derivative is to be understood as a functional derivative.  We define the Hamiltonian density $\mathcal{H}(q,p)$ as the Legendre transformation of the Lagrangian density with respect to the time derivatives of the fields (see equation \eqref{eq:def-ham-density}). One can then rewrite the action functional as in equation \eqref{eq:action-ham}.
\begin{align}
    S &= \int d^dx \,\, \mathcal{L}(q,\dot{q}) \label{eq:S-Lagrangian}\\
    p_n &:= \frac{\delta \mathcal{L}}{\delta \dot{q}^n} \label{eq:def-can-mom}\\
    \mathcal{H}(q,p) &:= \dot{q}^n p_n - \mathcal{L}(q,\dot{q}) \label{eq:def-ham-density}\\
    S &= \int d^dx \,\, \left( \dot{q}^n p_n - \mathcal{H}(q,p)  \right) \label{eq:action-ham}
\end{align}
Due to the Legendre transformation we have now found a function $\mathcal{H}(q,p)$ that depends on two sets of variables which we call the \emph{canonical variables} $q^n(x)$ and the \emph{canonical momenta} $p_n(x)$. We call the tuple $(q, p)$ the \emph{canonical coordinates} or \emph{phase space coordinates} and the space that they describe \emph{phase space}.\\ 

We can think of gauge theories as theories where, at any given time, the dynamical content of the theory is relative to an arbitrary reference frame \cite{Henneaux-Teitelboim}. Hence the general solution of a gauge theory necessarily contains arbitrary functions of time, since it is always permitted to transform the reference frame. This arbitrariness leads to the canonical coordinates not being completely independent and therefore it is equivalent to the existence of constraints on the phase space. In a gauge theory, we will always find constraints $\{\Phi_m(q,p)=0,\, \,\, m=1,\dots,M\}$ that only depend on the canonical coordinates and not on their time derivatives.\\

The first type of constraint follows directly from the definition of the canonical momenta (equation \eqref{eq:def-can-mom}), which is why we call them \emph{primary constraints}. The simplest primary constraint is given by a vanishing canonical momentum $\Phi(q,p)= p_n = 0$, however more complicated types are also common. Since we have not made use of the time evolution the primary constraints do not restrict the kinematics and are true identities.

A consequence of the existence of primary constraints is that the Hamiltonian becomes non-unique in phase space since we are free to add primary constraints with arbitrary coefficients $u^m(q,p)$ to the Hamiltonian. We call this function the \emph{total Hamiltonian} (see equation \eqref{eq:Ham-wth-primary}). The phase space hypersurface described by $\{\Phi_m = 0,\,\, \forall m\}$ is called the \emph{primary constraint surface} and on this surface the Hamiltonian is still uniquely defined. 
\begin{equation}
\label{eq:Ham-wth-primary}
    \mathcal{H}_T := \mathcal{H} + u^m(q,p) \,\,\Phi_m
\end{equation}
For two phase space functions $F,G$ we define the \emph{Poisson bracket} by equation \eqref{def:poisson}.

\begin{equation}\label{def:poisson}
    \{F, G\} :=  \frac{\delta F}{\delta q^n} \frac{\delta G}{\delta p_n} -\frac{\delta G}{\delta q^n} \frac{\delta F}{\delta p_n} 
\end{equation}
 Using the Poisson bracket we can write the time evolution of a phase space function $F$, that follows from the total Hamiltonian of equation \eqref{eq:Ham-wth-primary} as equation \eqref{eq:time-evolution-primary}. At this point this is the most general time evolution we can write down. 
 \begin{equation}\label{eq:time-evolution-primary}
     \dot{F} = \{F , H \} + u^m\{F, \Phi_m \}  
 \end{equation}
  Since we need the primary constraints to hold for all times --- in order for the formalism to be consistent --- we arrive at equation \eqref{eq:secondary-constraints}. If equation \eqref{eq:secondary-constraints} yields a relation that is independent of the arbitrary parameters $u^m$ and independent of the primary constraints, then we call it a \emph{secondary constraint}. Since we need to make sure it is also conserved in time we iterate this procedure for all secondary constraints and potentially end up with more constraints.\footnote{The consistency requirements that stem from secondary constraints are sometimes also referred to as tertiary, quaternary, etc. constraints. For reasonable physical theories the termination of this procedure is guaranteed.}
  \begin{equation}\label{eq:secondary-constraints}
        \dot{\Phi}_m = \{\Phi_m , H \} + u^{m'}\{\Phi_m, \Phi_{m'} \}  = 0
  \end{equation}
  The only difference between primary and secondary constraints is that secondary constraints do restrict the kinematics, since we have made use of the time evolution in order to find them. Note that this means that we are not allowed to add the secondary constraints to the Hamiltonian in the manner that we have done with the primary constraints in equation \eqref{eq:Ham-wth-primary}.
    Since further distinction between primary and secondary constraint is not needed we denote the complete set of constraints by $\{\Phi_j, \,\,\, j=1,\dots,J \}$.\\

    If we apply the time evolution equation \eqref{eq:time-evolution-primary} to the full set of constraints we will not find any new constraints. However the set of time evolution equations can be seen as a set of differential equations for the --- a priori --- arbitrary phase space functions $u^m(q,p)$. Solving this system we can write the general solution using the homogenous solution $V_a^m$ and a particular solution $U^m$ as in equation  \eqref{eq:general-sol-total-ham}. The coefficients of the homogeneous solution $v^a$ are then truly arbitrary. 
    \begin{equation} \label{eq:general-sol-total-ham}
            u^m = U^m + v^a\,\,V_a^m
    \end{equation} 
  We write the $\approx$ sign to indicate an equality that holds on the constraint surface, e.g. $\mathcal{H}_T \approx \mathcal{H}$, we call this relation \emph{weak equality}. The time evolution generated by the total Hamiltonian via $\dot{F} \approx \{F, \mathcal{H}_T \}$ is equivalent to the Lagrangian time evolution. \\
    
    An important property of the constraints is whether they are \emph{first class} (see equation \eqref{def:firstclass}) or \emph{second class} constraints (see equation \eqref{def:secondclass}).\footnote{Second class constraints require further treatment, since they do not occur in the theory being analysed in this work we again refer the reader to reference \cite{Henneaux-Teitelboim}.} Note that the definition of the class relies on the weak equality, meaning that first class constraints Poisson-commute with all other constraints up to terms proportional to constraints. This also implies that if all the constraints are first class the Poisson bracket algebra of the constraints closes automatically.
    \begin{align}
   \Phi_k\,\, \text{first class constraint}  \quad \Leftrightarrow  \forall j:   \{ \Phi_k,  \Phi_j \}&\approx 0  \label{def:firstclass} \\
   \Phi_k\,\, \text{second class constraint} \label{def:secondclass}  \quad \Leftrightarrow 
\exists j : \{ \Phi_k,  \Phi_j \}& \not\approx 0 
\end{align}
  First class constraints can be interpreted an generators of gauge transformations.\footnote{Mathematically this is only guaranteed for first class primary constraints, however for reasonable physical theories this is also true for first class secondary constraints. This is also known as the \emph{Dirac conjecture}. Counterexamples exist and are known. \cite{Henneaux-Teitelboim}} Gauge transformations $\delta F$ on a phase space function $F$ are generated as in equation \eqref{eq:gauge-transf-F}, where $\Phi_a := \Phi_m\,V^m_a $.
  \begin{equation}\label{eq:gauge-transf-F}
      \delta F = \{F, \Phi_a\} \, v^a
  \end{equation}
  We can add all first class constraints to the total Hamiltonian to describe the time evolution with the full gauge freedom accounted for. We call this the \emph{extended Hamiltonian}, see equation \eqref{def:total-Hamiltonian}, where $\{\gamma_a\}$ is the set of all first class constraints.
  \begin{equation}\label{def:total-Hamiltonian}
        \mathcal{H}_E := \mathcal{H}_T + u^a \gamma_a 
  \end{equation}
  The extended Hamiltonian generates time evolution via equation  \eqref{eq:time-evolution-extended-hamiltonian}. This can be seen as an extension of the Lagrangian framework since the full gauge freedom is now manifest in the time evolution. 
  \begin{equation}\label{eq:time-evolution-extended-hamiltonian}
      \dot{F}(q,p) \approx \{F, \mathcal{H}_E \}, \quad \Phi_j \approx 0
  \end{equation}
  
  This only concerns gauge variant quantities, since for gauge invariant quantities (\emph{observables}) all the Hamiltonian time evolutions are equivalent $ \mathcal{H}_E \Leftrightarrow \mathcal{H}_T \Leftrightarrow \mathcal{H} $.
  
  Furthermore it is a general feature of \emph{generally covariant} theories, that the Hamiltonian vanishes weakly $\mathcal{H}\approx0$ --- implying that the Hamiltonian consists of a linear combination of constraints. In particular general relativity and thus also all supergravity theories are generally covariant theories. In this form the time evolution is interpreted as a yet another gauge transformation.

\section{Poisson bracket relations} \label{sec:poisson-relations}

In this appendix we list some Poisson bracket relations that are intermediate results or otherwise useful for for computations.

\begin{equation}
     \{M^{MN}, \Pi^{PQ}(M) \} = \left(-M^{MP}\, M^{QN} - M^{MQ}\, M^{PN}\right) \delta^{(4)}(x-y)
\end{equation}
\begin{align}
    \int d^4x \, \phi(x) \, \{ -e\,R(x) , \Pi^n{}_a(e)(y)  \} = & +2e\,\phi\,\left( R^n{}_a - \frac{1}{2} R \,e_a{}^n \right) \nonumber\\
    & -2e \,\left( \nabla_a \nabla^n \phi-  e_a{}^n \nabla^k\nabla_k\phi \right) \label{sugra:eqn:poisson-Ricci-Pie}\\
    \{ e_b{}^k, \Pi^n{}_a(e)\} = &-e_a{}^k \, e_b{}^n  \\
    \{ e, \Pi^n{}_a(e)\} = &  + e \,e_a{}^n \\
  \left\{ \frac{1}{e}, \Pi^n{}_a(e) \right\} = & -\frac{e_a{}^n}{e}  \\
    \left\{ g_{kl}, \Pi^n{}_a(e) \right\} = &  +2\,\,\delta^n_{(k}\,e_{l)a}    \\
    \left\{ g^{kl}, \Pi^n{}_a(e) \right\} = &  -2\,\,g^{n(k}\,e_a{}^{l)}    
\end{align}

\begin{align}
\{ H_\text{Ham}[\phi], e \} = & + \frac{\phi \Pi(e)}{6} \\
\left\{ H_\text{Ham}[\phi], \frac{1}{e} \right\}  = & - \frac{\phi \Pi(e)}{6e^2}\\
\{ H_\text{Ham}[\phi], e_b{}^k \}  = & +\frac{\phi}{2e} \Pi_{ab}(e)\,\,e_a{}^k - \frac{\phi}{6e} \Pi(e)\,e_b{}^k\\
\{ H_\text{Ham}[\phi], g^{rm} \}  = & + \frac{\phi}{e}\,e^{a(r} \,\Pi^{m)}{}_a(e)\,  \, -\frac{\phi}{3e} \Pi(e)\, g^{rm}\\
\{ H_\text{Ham}[\phi], M^{MN} \}  = & + \frac{6}{e} \phi\,\Pi^{MN}(M)\\
\end{align}

\begin{equation}
    \{L[\gamma], e_a{}^k \} = + \gamma_{ab}\, e^{bk} 
\end{equation}

\section{Other useful formulae}
\label{appendix:other}
In this appendix we discuss some general formulae that are needed for the computations of the main sections. \\

From the definition of the (vielbein) determinant and the fact that its variation can be expressed as $\delta e = e\, e_n{}^a \,\delta e_a{}^n$ we find the following identity. 
\begin{align}
    e_a{}^n \, \partial_p e_n{}^a = e^{-1} \,\partial_p e = - e \, \partial_pe^{-1}
\end{align}

Furthermore we can exploit the fact that we can \emph{over-antisymmetrise} a tensor to make it vanish --- e.g. we take an object with D$+1$ indices in D dimensions and antisymmetrise them to get zero --- this is sometimes called the \emph{Schouten identity}. One can think of this identity as the fact that there cannot be D$+1$ linearly independent vectors in D dimensions.
Equation \eqref{eq:schouten-identity} states the identity in four dimensions, where $\epsilon$ is the Levi-Civita symbol in four dimension and $v^k$ are vector components.
\begin{equation}\label{eq:schouten-identity}
    \epsilon^{[lmnr}\,\, v^{k]} = 0
\end{equation}
If we expand this expression we arrive at equation \eqref{eq:schouten-identity-2}.
\begin{equation}\label{eq:schouten-identity-2}
    \epsilon^{lmnr}\,\, v^k = -4\, \epsilon^{k[lmn} \,\,v^{r]}
\end{equation}

\section{Treatment of coset constraints example SL(n)/SO(n) } \label{appendix:coset}
As explained in section \ref{sec:supergravity} we want to treat the E$_{6(6)}(\mathbb{R})/$USp$(8)$ coset constraints on $M_{MN}$ as being \emph{implicit}. To explain this formalism we consider the much simpler coset SL(n)/SO(n) as an example to demonstrate how one would explicitly or implicitly treat the coset constraints.\\

We start with the sigma model Lagrangian of a generic symmetric matrix $M_{MN}$ with $M,N=1,...,n$ and just the time derivative for simplicity. The inverse matrix is defined via $M_{MN}\, M^{NK} = \delta^K_M$. We then explicitly add the SL(n)/SO(n) coset constraint $c := \det(M) - 1$ with the Lagrange parameter $\phi$ to the Lagrangian arriving at equation \eqref{coset-constraints:Lagrangian}. This is the starting point for the explicit treatment of the coset constraints.  
\begin{equation}
 \label{coset-constraints:Lagrangian}
    \mathcal{L} = -\frac{1}{2} \dot{M}_{MN} \dot{M}^{MN} + c\, \phi
\end{equation}

We can now calculate the canonical momenta and find the primary constraint \eqref{coset-primary-constraint}.
\begin{align}
    \Pi^{MN}(M) &= -2 \dot{M}^{MN}\\
    \Pi(\phi) &= 0 \label{coset-primary-constraint}
\end{align}
The canonical Hamiltonian of this theory is given by equation \eqref{coset-hamilton}, where $\Pi_{MN}(M) := -\Pi^{RS}(M) \, M_{RM} \, M_{SN}$. The total Hamiltonian is given by $\mathcal{H}_T := \mathcal{H} + u \, \Pi(\phi)$. 
\begin{equation}\label{coset-hamilton}
    \mathcal{H} = +\frac{1}{8} \Pi_{MN}(M)\, \Pi^{MN}(M) + \dot{\phi}\,\Pi(\phi) -c\,\phi
\end{equation}
As explained in appendix \ref{sec:can-analysis-appendix} we now need to check the consistency of the primary constraint \eqref{coset-primary-constraint} using equation \eqref{eq:time-evolution-primary} and find that there is one secondary constraint given by equation \eqref{coset-secondary} as intended.
\begin{equation}\label{coset-secondary}
    c = \det(M) - 1 = 0
\end{equation}
Considering the equation \eqref{eq:time-evolution-primary} now for the consistency of the secondary constraint we find that there is one tertiary constraint given by equation \eqref{coset-tertiary}. In this calculation we have used equation \eqref{coset-identity-det} which follows directly from the fundamental Poisson bracket of a symmetric matrix with its canonical momentum \eqref{coset-fund-brack-sym} --- which is in form identical to the fundamental Poisson bracket in equation  \eqref{eqn:can-rel-Poisson-scalar}.
\begin{align}
     \det(M) \, M_{MN} \, \Pi^{MN}(M) &=: p = 0 \label{coset-tertiary}\\
    \{ \det(M) , \Pi^{MN}(M) \} &= \det(M) \, M^{MN} \label{coset-identity-det}\\
    \{M_{MN}(x), \Pi^{PQ}(M)(y) \} &= \left(\delta^P_M \, \delta^Q_N + \delta^P_N \,\delta^Q_M \right) \delta(x-y) \label{coset-fund-brack-sym}
\end{align}
The tertiary constraint p --- induced by the consistency of the coset constraint $\det(M)=1$ --- can be interpreted as a coset constraint on the momenta implying the tracelessness of the sl(n) algebra element $M_{RN} \, \Pi^{NS}(M)$.\\

Applying equation \eqref{eq:time-evolution-primary} to the tertiary constraint we do not find any new constraints and the constraint system is consistent.\\

We immediately find that the primary constraint $\Pi(\phi)$ is a first class constraint. The secondary and tertiary constraints however form a second class system of constraints since the right hand side of equation \eqref{coset-second-class} necessarily contains a constant term and cannot be rewritten solely in terms of the secondary constraint --- the factor $n$ here comes from $M_{MN} \, M^{MN} = \delta^M_M =n$. 
\begin{align}
    \{c,c\} &= 0 \\
    \{p,p\} &= 0 \\
    \{c,p\} &= n\, \det(M)^2 \label{coset-second-class}
\end{align}

The presence of these two second class constraints tells us that we should define a Dirac bracket $\{g,f\}_{DB}$ as defined in reference \cite{Henneaux-Teitelboim} given by equation \eqref{coset-dirac} where $f,g$ are two functions on phase space. We find by construction that all constraints commute in the Dirac bracket.
\begin{equation}\label{coset-dirac}
    \{f,g\}_{DB} := \{f,g\} - \frac{1}{n\,\det(M)^2} \left( \{f,p\} \{c,g\} - \{f,c\}\{p,g\}  \right)
\end{equation}

One may have noticed that equation \eqref{coset-identity-det} --- which is a general result for any symmetric matrix --- seems incompatible with the secondary constraint $\det(M)=1$ which defines the SL(n)/SO(n) coset. This seeming inconsistency is resolved by the fact that we are not allowed to apply the canonical constraints before evaluating all Poisson brackets fully.

For the Dirac bracket this is no longer true however \cite{Henneaux-Teitelboim} and we are free to apply the canonical constraints inside the Dirac bracket yielding the more expected equation \eqref{coset-dirac-det}. 
\begin{equation}\label{coset-dirac-det}
     \{ \det(M) , \Pi^{MN}(M) \}_{DB} = 0
\end{equation}

We can furthermore compute the number of degrees of freedom. We have the canonical variables $M_{MN}, \,\Pi^{KL}(M),\, \phi,\,\Pi(\phi)$ which have $\frac{n(n+1)}{2} + \frac{n(n+1)}{2} + 1 + 1$ components --- a priori the the variables $M_{MN}, \,\Pi^{KL}(M)$ are still generic symmetric matrices. We have one first class primary, one second class secondary and one second class tertiary constraints which means we should count them as $2\cdot1 + 1 + 1$ \cite{Henneaux-Teitelboim}. This leaves us with $n^2+n-2$ physical degrees of freedom in phase space or $\frac{n^2}{2} + \frac{n}{2} -1$ physical degrees of freedom in field space.

This is the full explicit treatment of the coset constraints in the SL(n)/SO(n) case. \\

The implicit treatment of the coset constraints uses the Lagrangian without adding the coset constraints explicitly $\mathcal{L} = -\frac{1}{2} \dot{M}_{MN} \dot{M}^{MN}$. We can nonetheless apply them after all Poisson brackets have been computed, thus considering them implicitly. The price one pays for the implicit treatment of the constraints is that we are not able to apply canonical constraints inside the brackets. In return we are able to skip the canonical analysis of the coset constraints --- which are much more complicated for the E$_{6(6)}(\mathbb{R})/$USp$(8)$ coset --- if we are not interested in applying them inside brackets. We can compute the canonical degrees of freedom by looking at the dimension of the coset, in this case $\dim($SL(n)/SO(n)$)=\frac{n^2}{2} + \frac{n}{2} -1$. We thus arrive at the same result as in the explicit treatment. \\

For an alternative canonical treatment of coset space sigma models using the vielbein formalism see reference \cite{Matschull:1994vi}.

\acknowledgments

This project has received funding from the European Research Council 
(ERC) under the European Union's Horizon 2020 research and innovation 
programme ("Exceptional Quantum Gravity", grant agreement No 740209).\\

I am supported by the  International  Max  Planck  Research  School  for  Mathematical and Physical Aspects of Gravitation, Cosmology and Quantum Field Theory. \\

 I am particularly grateful to Axel Kleinschmidt for his support, help and many useful discussions.
 Furthermore I would like to thank Matteo Broccoli, Franz Ciceri, Jan Gerken, Emanuel Malek, Hermann Nicolai and Stefan Theisen for comments and useful discussions.

\paragraph{Data Availability} Data sharing is not applicable to this article as no new data were created or analyzed in this study.

\providecommand{\href}[2]{#2}\begingroup\raggedright\endgroup


\begin{thebibliography}{10}

\bibitem{Cremmer:1979up}
E.~Cremmer and B.~Julia, {\it {The SO(8) Supergravity}},  {\em Nucl. Phys. B}
  {\bf 159} (1979) 141--212.

\bibitem{Cremmer}
E.~Cremmer, {\it Supergravities in 5 dimensions},  in {\em Superspace and
  Supergravity, Proceedings of Nuffield workshop in Cambridge (UK), June 16 -
  July 12, 1980} (S.~Hawking and M.~Rocek, eds.), Cambridge University Press,
  8, 1980.

\bibitem{cartan}
E.~Cartan, {\em Sur la structure des groupes de transformations finis et
  continus}.
\newblock PhD thesis, Paris, 1894.
\newblock (II Edition, 1933).

\bibitem{greenbook-lie-alg}
E.~de~Kerf, G.~B\"auerle, and A.~ten Kroode, {\em Lie Algebras, Finite and
  Infinite Dimensional Lie Algebras and Applications in Physics}.
\newblock Studies in mathematical physics. North Holland, 1997.

\bibitem{Hohm_2013First}
O.~Hohm and H.~Samtleben, {\it Exceptional form of d=11 supergravity},  {\em
  Physical Review Letters} {\bf 111} (Dec, 2013).

\bibitem{EFTI-E6}
O.~Hohm and H.~Samtleben, {\it {Exceptional field theory. I.
  E$_{6(6)}$-covariant form of M-theory and type IIB}},  {\em Phys. Rev. D}
  {\bf 89} (3, 2014) 066016.

\bibitem{Baguet:2015xha}
A.~Baguet, O.~Hohm, and H.~Samtleben, {\it {E$_{6(6)}$ Exceptional Field
  Theory: Review and Embedding of Type IIB}},  {\em PoS} {\bf CORFU2014} (2015)
  133, [\href{http://arxiv.org/abs/1506.01065}{{\tt arXiv:1506.01065}}].

\bibitem{Musaev:2014lna}
E.~Musaev and H.~Samtleben, {\it {Fermions and supersymmetry in E$_{6(6)}$
  exceptional field theory}},  {\em JHEP} {\bf 03} (2015) 027,
  [\href{http://arxiv.org/abs/1412.7286}{{\tt arXiv:1412.7286}}].

\bibitem{berman2020geometry}
D.~S. Berman and C.~Blair, {\it The geometry, branes and applications of
  exceptional field theory},  {\em International Journal of Modern Physics A}
  {\bf 35} (Oct, 2020) 2030014.

\bibitem{Hohm:2014qga}
O.~Hohm and H.~Samtleben, {\it {Consistent Kaluza-Klein Truncations via
  Exceptional Field Theory}},  {\em JHEP} {\bf 01} (2015) 131,
  [\href{http://arxiv.org/abs/1410.8145}{{\tt arXiv:1410.8145}}].

\bibitem{Bossard:2015foa}
G.~Bossard and A.~Kleinschmidt, {\it {Loops in exceptional field theory}},
  {\em JHEP} {\bf 01} (2016) 164, [\href{http://arxiv.org/abs/1510.07859}{{\tt
  arXiv:1510.07859}}].

\bibitem{PhysRevLett.57.2244}
A.~Ashtekar, {\it New variables for classical and quantum gravity},  {\em Phys.
  Rev. Lett.} {\bf 57} (Nov, 1986) 2244--2247.

\bibitem{PhysRevD.36.1587}
A.~Ashtekar, {\it New hamiltonian formulation of general relativity},  {\em
  Phys. Rev. D} {\bf 36} (Sep, 1987) 1587--1602.

\bibitem{Godazgar_2014Ashtekar}
H.~Godazgar, M.~Godazgar, and H.~Nicolai, {\it Einstein-cartan calculus for
  exceptional geometry},  {\em Journal of High Energy Physics} {\bf 2014} (Jun,
  2014).

\bibitem{dirac-lectures-QM}
P.~A.~M. Dirac, {\em Lectures on Quantum Mechanics}.
\newblock Belfer Graduate School of Science, Yeshiva University, 1964.

\bibitem{Henneaux-Teitelboim}
M.~Henneaux and C.~Teitelboim, {\em Quantization of gauge systems}.
\newblock Princeton University Press, 1992.

\bibitem{1962ADM}
R.~L. Arnowitt, S.~Deser, and C.~W. Misner, {\it {The Dynamics of general
  relativity}},  {\em Gen. Rel. Grav.} {\bf 40} (2008) 1997--2027,
  [\href{http://arxiv.org/abs/gr-qc/0405109}{{\tt gr-qc/0405109}}].

\bibitem{MTW}
C.~W. Misner, K.~S. Thorne, and J.~A. Wheeler, {\em Gravitation}.
\newblock W. H. Freeman, 1973.

\bibitem{Wald:1984}
R.~M. Wald, {\em {General Relativity}}.
\newblock Chicago Univ. Pr., Chicago, USA, 1984.

\bibitem{Nicolai:1992xx}
H.~Nicolai and H.-J. Matschull, {\it {Aspects of canonical gravity and
  supergravity}},  {\em J. Geom. Phys.} {\bf 11} (1993) 15--62.

\bibitem{Freedman:2012zz}
D.~Z. Freedman and A.~Van~Proeyen, {\em {Supergravity}}.
\newblock Cambridge Univ. Press, Cambridge, UK, 5, 2012.

\bibitem{HenneauxG2}
M.~Henneaux, A.~Kleinschmidt, and V.~Lekeu, {\it {Enhancement of hidden
  symmetries and Chern-Simons couplings}},  {\em Rom. J. Phys.} {\bf 61}
  (2016), no.~1-2 167, [\href{http://arxiv.org/abs/1505.07355}{{\tt
  arXiv:1505.07355}}].

\bibitem{dewittCanGR}
B.~S. DeWitt, {\it {Quantum Theory of Gravity. I. The Canonical Theory}},  {\em
  Phys. Rev.} {\bf 160} (1967) 1113--1148.

\bibitem{HOJMAN197688}
S.~A. Hojman, K.~Kucha\v{r}, and C.~Teitelboim, {\it Geometrodynamics
  regained},  {\em Annals of Physics} {\bf 96} (1976), no.~1 88 -- 135.

\bibitem{Kiefer:2004gr}
C.~Kiefer, {\em {Quantum gravity}}.
\newblock International Series of Monographs on Physics. Oxford University
  Press, 2012.

\bibitem{PhysRevD.33.2801}
A.~H. Diaz, {\it Hamiltonian formulation of eleven-dimensional supergravity},
  {\em Phys. Rev. D} {\bf 33} (May, 1986) 2801--2808.

\bibitem{PhysRevD.33.2809}
A.~H. Diaz, {\it Constraint algebra in eleven-dimensional supergravity},  {\em
  Phys. Rev. D} {\bf 33} (May, 1986) 2809--2812.

\bibitem{Strathdee:1986jr}
J.~Strathdee, {\it {Extended Poincare Supersymmetry}},  {\em Int. J. Mod. Phys.
  A} {\bf 2} (1987) 273.

\bibitem{Cremmer:1978km}
E.~Cremmer, B.~Julia, and J.~Scherk, {\it {Supergravity Theory in
  Eleven-Dimensions}},  {\em Phys. Lett. B} {\bf 76} (1978) 409--412.

\bibitem{Matschull:1994vi}
H.~Matschull and H.~Nicolai, {\it {Canonical treatment of coset space sigma
  models}},  {\em Int. J. Mod. Phys. D} {\bf 3} (1994) 81--91.

\end{thebibliography}
\end{document}